\documentclass[10pt]{iopart}

\expandafter\let\csname equation*\endcsname\relax
\expandafter\let\csname endequation*\endcsname\relax
\usepackage{graphics}
\usepackage{physics}
\usepackage{xcolor}
\usepackage{enumerate}
\usepackage{amsmath}
\usepackage{amssymb}
\usepackage{graphicx}
\usepackage{float}

\usepackage{hyperref}

\newcommand{\ie}{\textit{i.e.} }
\newcommand{\eg}{\textit{e.g.} }

\newcommand{\ATT}{ANTITER-IV}

\usepackage{iopams}

\begin{document}

\title[Minimization of the edge modes and near fields of a Travelling Wave Array antennas]{Minimization of the edge modes and near fields of a Travelling Wave Array antenna for WEST}

\author{V. Maquet$^1$ \& R. Ragona$^2$, F. Durodié$^1$, J. Hillairet$^3$}

\address{$^1$Laboratory for Plasma Physics, LPP-ERM/KMS, 1000 Brussels, Belgium\\
$^2$Technical University of Denmark, Department of Physics, 2800 Lyngby, Denmark\\
$^3$CEA, IRFM, F-13108 Saint-Paul-lez-Durance}
\ead{Vincent.Maquet@ulb.be \& ricrag@fysik.dtu.dk}
\vspace{10pt}
\begin{indented}
\item[]January 2023
\end{indented}

\begin{abstract}
The Travelling Wave Array (TWA) antenna was proposed as a promising alternative for Ion Cyclotron Resonant Heating (ICRH) in future fusion reactors. In this study, the possibility to make a TWA compatible with a tungsten environment like the WEST tokamak is assessed. For this purpose, two aspects of the antenna are investigated: the power spectrum and the near fields excited by the antenna. The sensitivity of these parameters to load and capacitor layout variations is taken into account while satisfying a proper antenna frequency response. The sensitivity of the power spectrum to frequency variation is also investigated to allow the possibility of fast feedback of the power deposition into the core plasma of WEST. The high resilience of the TWA to these variations is demonstrated and the main parameters of the TWA expected in WEST are compared to the WEST Q2 antenna for the same loading. Possible optimizations of the TWA antenna are discussed. The present study is fully transferable to a fusion reactor like DEMO or ARC.
\end{abstract}

\vspace{2pc}
\\

%
%
%

\section{Introduction}
The Travelling Wave Array (TWA) consists of an array of straps carrying a travelling wave along its structure. A TWA was proposed in the Ion Cyclotron Range of Frequencies (ICRF) as an alternative for fusion reactors to the common tightly packed and individually fed antenna straps used in present-day tokamaks (\eg AUG, WEST, JET, EAST, Alcator C-Mod) and foreseen for ITER to heat the plasma.

The TWA has been successfully tested on several devices for Fast Wave Current Drive (FWCD) \cite{Ogawa_2001,Takase_2001,Takase_2004}, Lower Hybrid (LH) launch \cite{jo2018coupling, takase2015plasma} and helicon waves launch \cite{Pinsker_2018,WI201867} as well as plasma production \cite{Ogawa_2001} and plasma start-up \cite{takase2015plasma}. This type of antenna presents several advantages:
\begin{enumerate}
    \item TWAs' power spectrum have a narrow toroidal extent, leading to directional fast waves and enabling localized power deposition into the core plasma \cite{Ogawa_2001,RAGONA2019854}.
    \item TWAs were shown to maintain a good impedance matching without dynamic tuning \cite{ikezi1997traveling,Ogawa_2001} even during Edge Localized Modes (ELMs).
    \item TWAs' power spectrum were also shown to be resilient to plasma loading variations \cite{GA_paper}.
    \item TWAs are fed only on each end of the array, reducing the number of coaxial lines and feeders needed compared to the individually fed straps antennas used in most machines in the ICRF, reducing the mechanical complexity and the overall system costs.
    \item TWAs use a larger number of straps compared to the conventional antennas to reduce the power density and therefore the voltage and currents on straps and the maximum fields excited near the antenna \cite{Ragona_2016}.
    \item The  remaining power at the end of the array which has not been coupled can be recirculated by means of a resonant ring scheme \cite{GA_paper,Ragona_2016}.
\end{enumerate}

A TWA was already proposed for ICRF heating in the 90s by the DIII-D team in the US but never implemented. More recently, the TWA concept in the ICRF was proposed for the future fusion reactor DEMO \cite{Ragona_2019,Van_Eester_2019} with a possible proof of concept in WEST \cite{RAGONA2019854}. The test on WEST would allow an assessment of the TWA in a tokamak with a reactor-relevant metallic environment and direct comparison with in-port ICRH antennas.

Similarly to other ICRF antenna concepts, the TWA should aim at minimizing the additional impurity influx induced during ICRF operations. Numerous physical phenomena can take place during ICRF antenna operation (\ie sheath, ponderomotive effects, edge resonances, etc.) and no model is yet able to give a complete picture of the physics taking place near and far from the active antenna. However, thanks to the experience gathered over years of experimental campaigns and numerical modelling \cite{Colas2021,BOBKOV2019131}, several guidelines for the antenna design have emerged:
\begin{enumerate}
    \item The antenna should minimize its ``near fields'' and more precisely the RF image currents and the electric field parallel to the magnetic field on plasma facing components (PFCs) near the antenna and protruding from the antenna frame such as limiters. This guideline aims at decreasing RF sheaths and related convective cells and led to the successful operation of a 3-strap ICRH antenna with W limiters in ASDEX Upgrade (AUG) \cite{Bobkov_2016,BOBKOV2019131}. The method used in AUG to minimize near fields was reproduced with the Alcator C-Mod field-aligned (FA) antenna \cite{lin_wright_wukitch_2020} and more recently with the JET A2 antennas. In a more general way, the other field components around the antenna should be as low as possible as they can lead to other non-linear phenomena like ponderomotive force \cite{van2015crude}.
    \item The antenna wave spectrum should maximize the power absorption into the core plasma which requires exciting a precise range of $k_\parallel$ wavenumber and avoid as much as possible low-$\abs{k_\parallel}$ spurious excitation in the wave spectrum. This first guideline should mainly reduce far-field effects due to non-absorbed RF power \cite{kohno2015numerical}. In addition, in the presence of low densities in front of the antenna ($n_e<10^{17}$ m$^{-3}$), this guideline also reduces power loss and edge modes excited due to the presence of an LH resonance in the antenna edge density profile and the spurious antenna coupling to slow waves due to a FS misalignment with the background magnetic field \cite{Messiaen_2021,maquet_2021}. Moreover, a good correlation was found between ICRH-specific impurity generation and the low toroidal part of the power spectrum excited by ICRH antennas used in experiments minimizing near fields \cite{maquet_messiaen_2020}.
\end{enumerate}

This paper presents the following structure. Section \ref{sec:TWA_on_West} and \ref{sec:model_description} describe the TWA layout presently foreseen for WEST and describe the model and experimental parameters considered in the rest of the paper. Section \ref{sec:ATT} assess the possibility to create an optimal TWA power spectrum depleted from low-$k_\parallel$ excitation with \ATT{}. In section \ref{sec:HFSS}, the power spectrum and the fields on the antenna PFCs are minimized using the 3D High-Frequency Simulation Software (ANSYS HFSS) with a flat antenna geometry including the antenna limiters. The fields and the power spectra of the TWA are compared with the WEST Q2 antenna. The sensitivity of the solution to capacitance, loading and frequency variations is discussed. In section \ref{sec:Optimization}, possible optimization are discussed. Finally, conclusions are drawn.

\section{A TWA on WEST}\label{sec:TWA_on_West}
\begin{figure*}[t]
    \centering
    \includegraphics[width=0.9\textwidth]{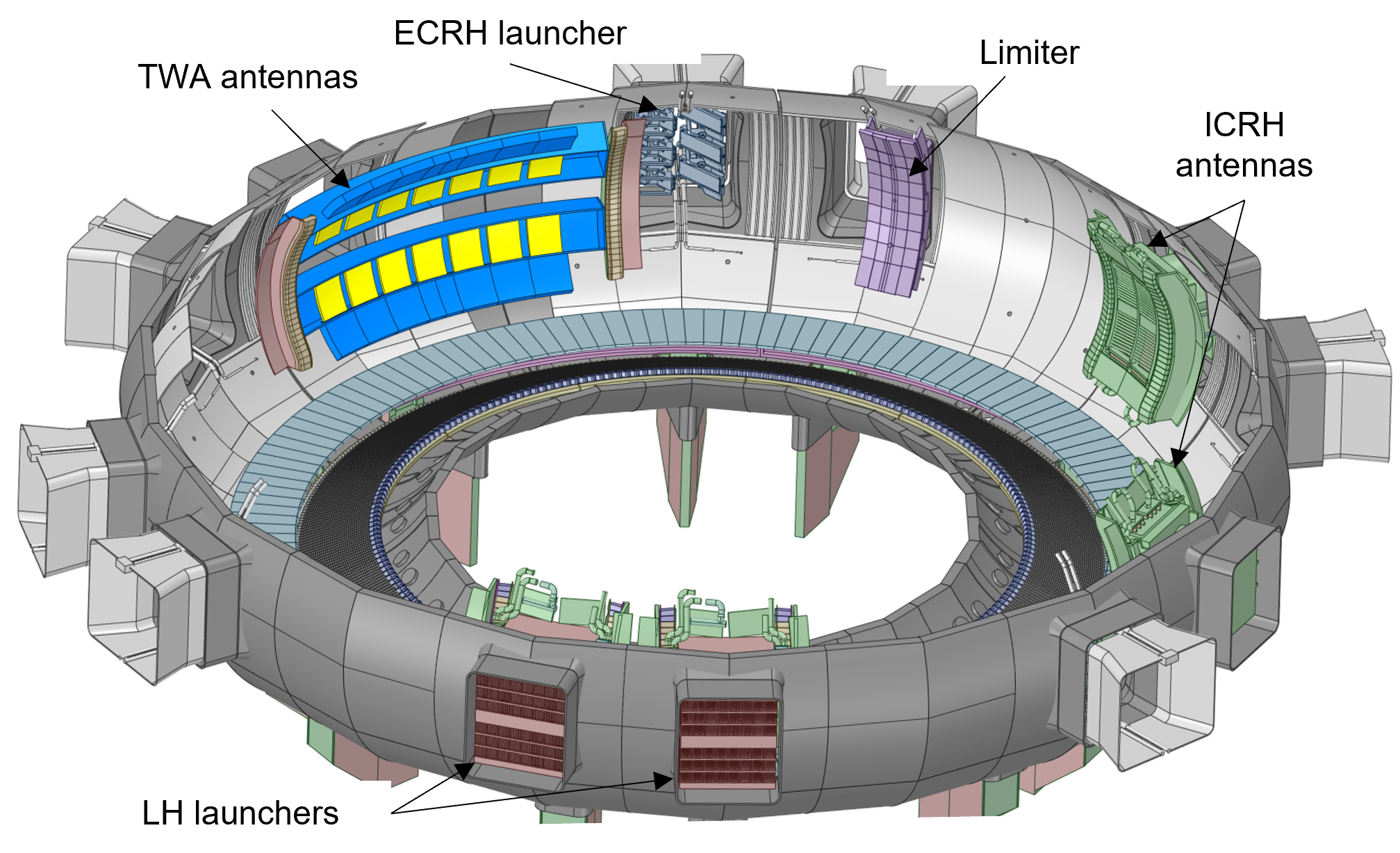}
    \caption{Overview (horizontal cut) of the possible layout of WEST showing the additional heating systems: electron cyclotron (ECRH), ion cyclotron (ICRH), lower hybrid (LH) within their allocated ports.}
    \label{fig:WEST_outlook}
\end{figure*}

An overview of the possible layout of WEST showing the auxiliary heating systems is presented in the figure \ref{fig:WEST_outlook}.
The proposed WEST TWA system is composed of two stacked rows of 7 straps recessed by 6 cm inside the antenna box. In the model studied, the strap recess was chosen to be 2 cm greater than the Q2 ICRH antennas to demonstrate the good coupling performance expected with a TWA despite a larger evanescent layer anticipated in future fusion reactors. Two poloidal bumpers are placed on each side of the structure and define the total toroidal space used by the system. The limiters are placed 2 cm in front of the antenna aperture. The system is located in between sectors Q3B (ECRH) and Q4B (diagnostics) and uses the equatorial port Q4A as unique access. The main WEST parameters are described in \cite{bourdelle2015west} and are reported here in table \ref{tab:WEST_main_par} for convenience. The antenna foreseen would be entirely made of stainless steel (SS) and actively water-cooled. The system power requirements are 3 MW -- 30 s and 1 MW -- 1000 s.

In 2021, an ICRF TWA mock-up was tested in TITAN at high power \cite{Ragona_2022}. This test was performed to verify the possibility to compensate for antenna uncertainties during manufacturing and to characterize the antenna sensitivity to thermal deformations. The antenna RF conditioning did not show any sign of arcs and tripping of the generators. In particular, successful tests at 500 kW / 60 s, 1 MW / 20 s, 1.75 MW / 5 s and 2 MW / 3 s confirmed the properties of the SS mock-up and paved the way for a WEST TWA antenna.  

\begin{table}[h]
    \centering
    \caption{Main WEST parameters. \label{tab:WEST_main_par}}
    \begin{tabular}{l r|l }
     Major radius & $R_0$ (m)& 2.5\\ 
     Minor radius & $a$ (m)& 0.5\\  
     Magnetic field & $B_0$ (T)& 3.7    
    \end{tabular}
\end{table}

\section{Model Description}\label{sec:model_description}
\begin{figure}[th]
    \centering
    \includegraphics[width=0.95\linewidth]{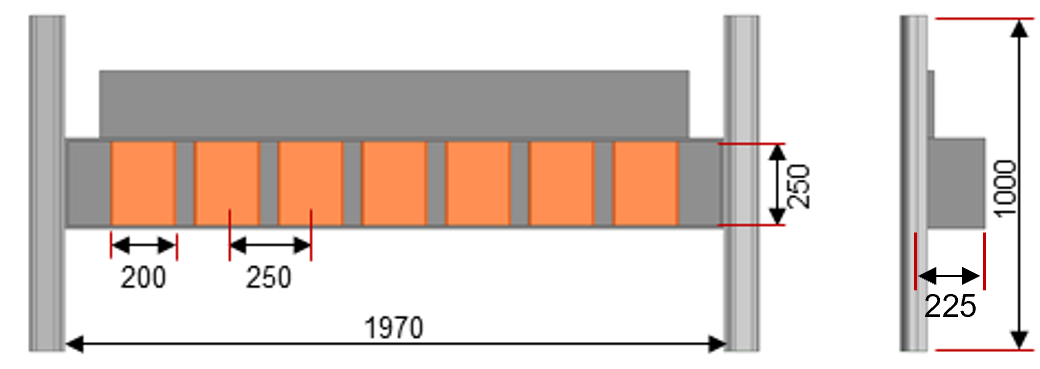}
    \caption{TWA model used in the simulation with main dimensions expressed in mm.}
    \label{fig:TWA_sketch}
\end{figure}

An equivalent flat antenna model is used for the modelling. The current model could be upgraded to a more realistic curved model in future studies. The main dimensions of one TWA row are shown in figure \ref{fig:TWA_sketch}. Two limiters are added to the HFSS model to properly evaluate the spectra and the near fields around the antenna. The region inside the antenna box is modelled as vacuum. The external region that models the plasma is described in detail in the following section.

\begin{figure}[th]
    \centering
    \includegraphics[width=0.95\linewidth]{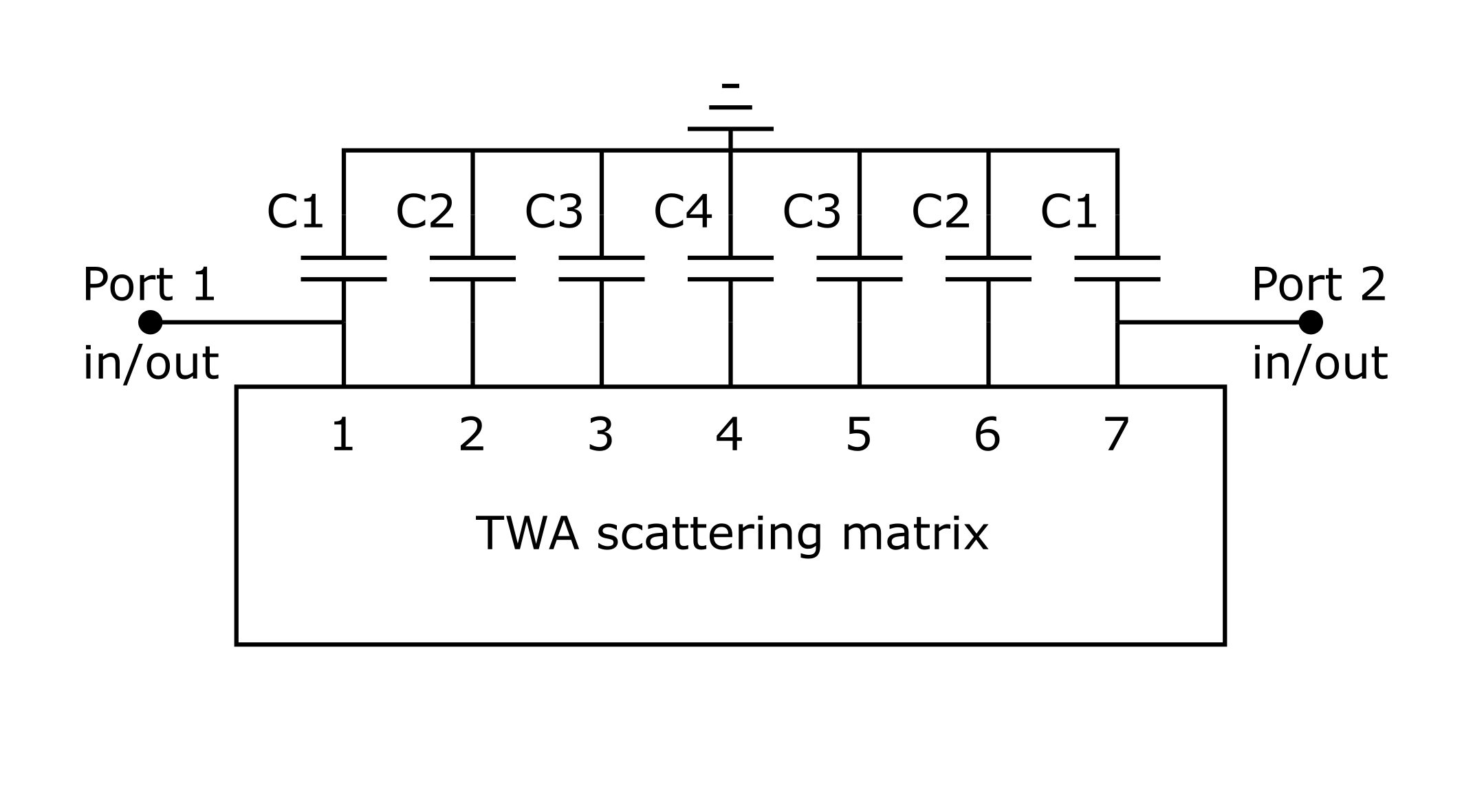}
    \caption{Schematic of the TWA circuit built from the scattering matrix calculated in the simulations. The resulting 2-port structure is obtained by adding a capacitor to each strap and two input/output (external) ports.}
    \label{fig:TWA_circuit_schematic}
\end{figure}

In the HFSS model, each strap is connected to a wave-port inside its respective capacitor box. For both \ATT{} and HFSS, a scattering ($S$) matrix of the system is calculated. The 7-port S-matrix is used to build a circuit model of the TWA. One capacitor is added to each port. On the first and last straps, a port is connected in parallel to provide the input/output connections thus reducing the system to a 2-port structure, as shown in figure \ref{fig:TWA_circuit_schematic}.
In the real antenna, each strap is tuned by a top capacitor housed in an independent box. A parallel plate capacitor configuration design is used where the capacitance is produced directly by the gap between the straps and the box housing it, so only metallic components are used under the machine vacuum. In principle, the 7 capacitors can have an arbitrary capacitance value. However, their layout is kept symmetric to enable co- and counter operations of the antenna, \ie input and output can be swapped without a change in the antenna response.

\begin{figure}[h]
    \centering
    \includegraphics[width=0.95\linewidth]{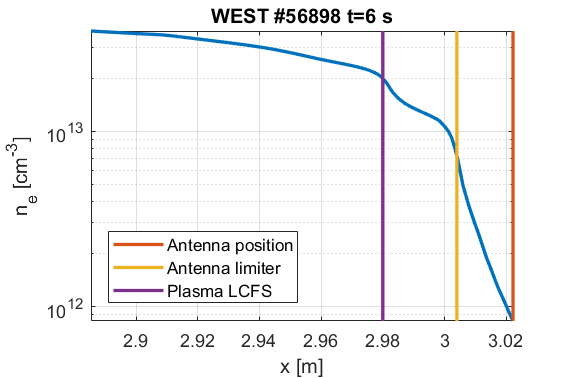}
    \caption{Electron density profile WEST shot \#56898 at 6 s along with the antenna front face, the antenna limiter and the LCFS positions.}
    \label{fig:dens_profile}
\end{figure}

The plasma parameters used for the \ATT{} and TOPICA simulations are extracted from the WEST pulse \#56898. In the experiment, the plasma composition is D-(H) with a minority concentration of around 4\%. The electron density profile is reconstructed from reflectometry data and is shown in figure \ref{fig:dens_profile} for $t=6$ s. The antenna front face, the limiters and the LCFS positions are respectively 3.022 m, 3.004 m and 2.980 m. As the reflectometer is at a different toroidal location than the antenna, the profile is not measured in front of the antenna. Consequently, the radial location of the profile is subject to uncertainties coming from the magnetic equilibrium reconstruction and from the noise in the reflectometer data.

\section{Wave Spectrum analysis with \ATT{}} \label{sec:ATT}
This section assesses the feasibility to have a TWA power spectrum depleted from low toroidal wavenumbers using \ATT{}. 

In this paper, the power spectrum is defined as the predominant FW Poynting components ($E_y\cross H_z^*$) radiated at the antenna aperture
\begin{align}
    2 P_{rad}(k_y,k_z)= \Re{E_y(k_y,k_z)\cross H_z(k_y,k_z)^*}.
\end{align}
The power spectra $P_{rad}(k_z,k_y)$ presented in the paper are normalized such that the total power radiated by the antenna $P_{tot}$ is 1 W,
\begin{align}
    P_{rad,tot}=\int P_{rad}(k_y,k_z) \dd k_z \dd k_y =1 \text{ W}.
\end{align}
The ratio of low-$\abs{k_z}$ power spectrum is defined for toroidal wavenumbers $\abs{k_z}<1.5 k_0$ as $\rho=P_{rad,z}(\abs{k_z}<1.5 k_0)/P_{rad,tot}$. As already discussed before, $\rho$ accounts for the part of the power which is not correctly absorbed in the plasma centre \cite{maquet_messiaen_2020,maquet_2021}.

The WEST plasma profile presented in figure \ref{fig:dens_profile} shows high coupling characteristics, the LH resonance is absent and the possible slow waves excited by the antenna are evanescent. This high coupling profile does not seem to correlate with the low radiation resistance measured in the experiment. This difference could be attributed to magnetic reconstruction errors or geometric effects. To take this uncertainty into account, the density profile in front of the antenna in \ATT{} is shifted away by 3 cm to reproduce the radiation resistance measured at the Q2 ICRH antenna at t=6 s in WEST shot \#56898. The choice of this particular loading in ANTITER does not hamper the conclusion of the following section.

\subsection{\ATT{} results} \label{subsec:ATT}
The possibility to shape the power spectrum of a TWA is explored with \ATT{}, using the WEST parameters mentioned above. The low-$\abs{k_z}$ spectrum excited by the antenna is not correctly absorbed in the plasma core and can lead to deleterious edge power losses as well as undesirable modes in the presence of a LH resonance in the edge plasma density profile \cite{maquet_messiaen_2020, Messiaen_2021, maquet_2021}. 

In ICRF antennas with more than two straps, one can actively minimize this lower part of the power spectrum by changing the current distribution in both amplitude and phase \cite{Messiaen_2020, maquet_messiaen_2020}.
In the TWA case, the wave spectrum launched by the antenna is set by the antenna dimensions, its tuning capacitor layout and by the antenna loading and frequency. For a given antenna geometry, these three last factors modify the current distribution on the strap array, therefore changing its power spectrum. Up to now, the capacitor layout was chosen to maximize the bandpass of the TWA for a given loading. However, this leads to power spectra which have a spurious excitation of low toroidal wavenumbers.

\begin{figure}
    \centering
    \includegraphics[width=0.95\linewidth]{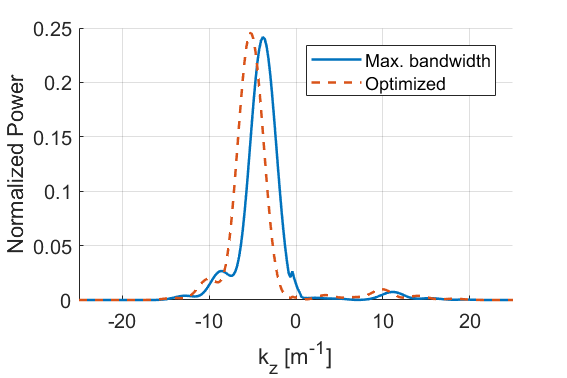}
    \caption{Typical normalized power spectrum found with \ATT{} for a capacitor layout optimizing the bandwidth of the antenna and one optimizing the power spectrum.}
    \label{fig:ATT_TWA_pwr_spct_nocoax}
\end{figure}

\begin{figure}
    \centering
    \includegraphics[width=0.95\linewidth]{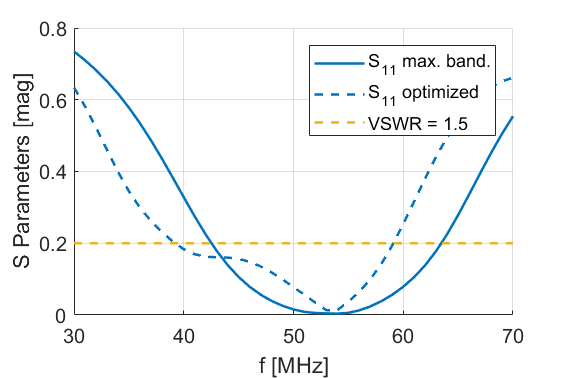}
    \caption{Antenna frequency response found for a capacitor layout optimizing the bandwidth of the antenna and one optimizing the power spectrum. The maximum VSWR limit that the WEST generator can handle is also apparent.}
    \label{fig:ATT_TWA_BW}
\end{figure}

Another possibility is to optimize the capacitor layout of the TWA to minimize the low toroidal part of the power spectrum in \ATT{}. Changing the capacitor layout also changes the TWA frequency response. Therefore, another constraint is needed to enforce a sufficient bandpass around the ICRH heating frequency of interest. Using a simplex optimization, it is possible to considerably reduce the unwanted low-$\abs{k_z}$ excitation of the power spectrum. This result is presented in figure \ref{fig:ATT_TWA_pwr_spct_nocoax} where the antenna power spectrum for a capacitor layout maximizing the frequency bandwidth and one minimizing the low-$\abs{k_z}$ excitation. The impact of the optimization on the bandwidth is presented in figure \ref{fig:ATT_TWA_BW}. The optimization shifts the frequency response of the antenna to lower frequencies and makes it narrower. 

While \ATT{} can formally treat the plasma, it cannot take into account the complex geometry of a real antenna. This comprises the finite thickness of the antenna straps as well as PFCs like the limiters and the non-ideal FS description. The complex geometry gives the possibility to estimate the amplitude of the fields taking place near the antenna PFCs like the limiters and the FS. These details can play an important role in the antenna's performance during its operation in front of a plasma. For these reasons, the antenna will be further analyzed using HFSS in front of an equivalent dielectric in this manuscript and should be validated in front of a plasma load in later works.

\section{HFSS fields and wave spectrum analysis} \label{sec:HFSS}
The result from \ATT{} suggests the possibility to obtain a power spectrum depleted of low toroidal excitations. This possibility is further investigated by using HFSS which gives the opportunity to extend the analysis to better characterize the antenna near fields. This analysis proved to be successful in reducing impurity release when implemented in AUG \cite{BOBKOV2019131}.

HFSS allows the use of complex tensors to describe an arbitrary material. However, the computational cost of such a method is excessive when a cold plasma tensor is used and might suffer from ill-defined boundary conditions. In our model, the plasma is therefore replaced by an equivalent dielectric load. The relative permittivity ($K_\mathrm{D}$) of the material is defined by a profile which is a function of the (tokamak) radial direction. This profile is calculated as $K_\mathrm{D}=\mathrm{S}-\mathrm{D}$ \cite{messiaen2011ppcf}, where $\mathrm{S}$ and $\mathrm{D}$ are the Stix dielectric tensor components \cite{stix1992waves}, and aligns the radial cutoff positions of each wavenumber $k_z$ in the dielectric with the plasma ones for $k_y=0$.

\subsection{Choice of the equivalent dielectric load}\label{subsec:equivalent_dielectric}
The dielectric load put in front of the antenna was chosen to reproduce the radiation resistance measured on the Q2 ICRH antenna in shot \#56898 at t=6 s. A flat model of the Q2 antenna was used in front of a dielectric profile $K_\mathrm{D}(x)$. The dielectric profile was then cut at different positions. 

In parallel, an equivalent analysis was performed with TOPICA. Good correspondence was found between the numerical and experimental values of the radiation resistance $R_c$ with a profile cut at the LCFS, for both HFSS and TOPICA. The same profile was then used in front of the TWA.

\begin{figure*}[h]
\begin{minipage}{0.49\linewidth}
    \centering
    \includegraphics[width=\linewidth]{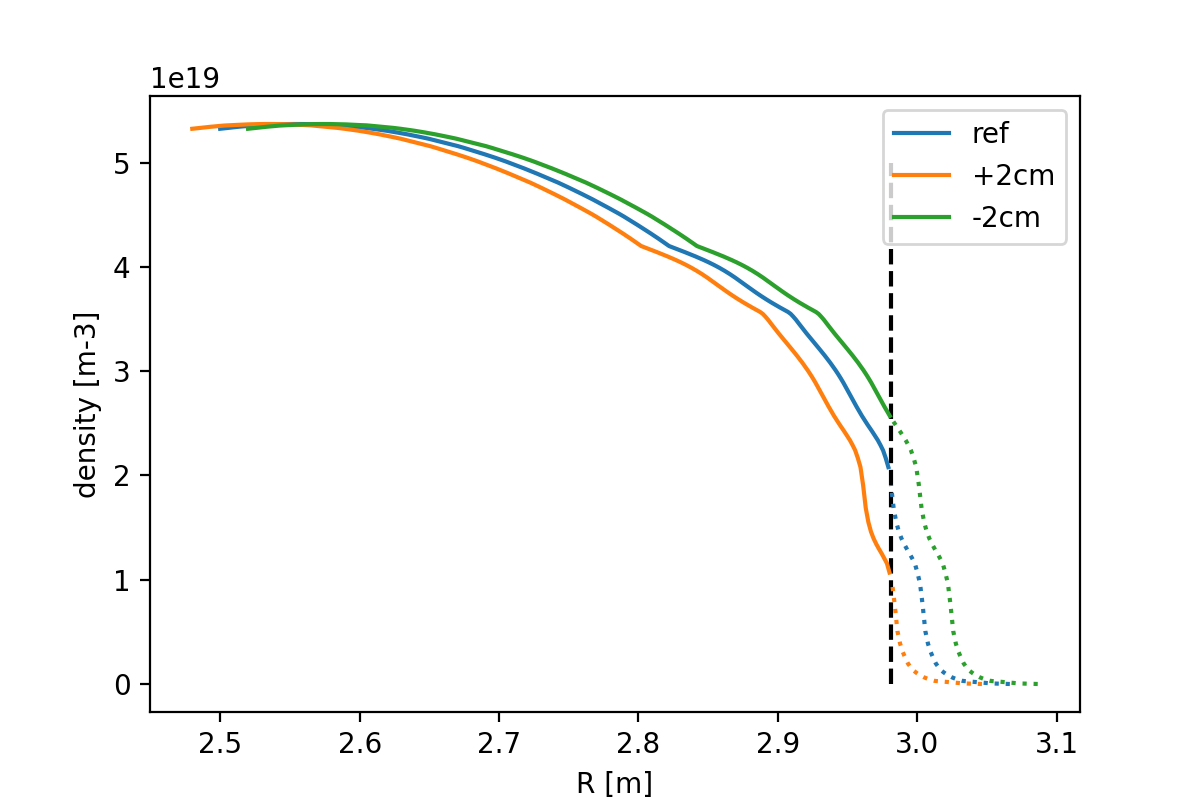}
\end{minipage}
\begin{minipage}{0.49\linewidth}
    \centering
    \includegraphics[width=\linewidth]{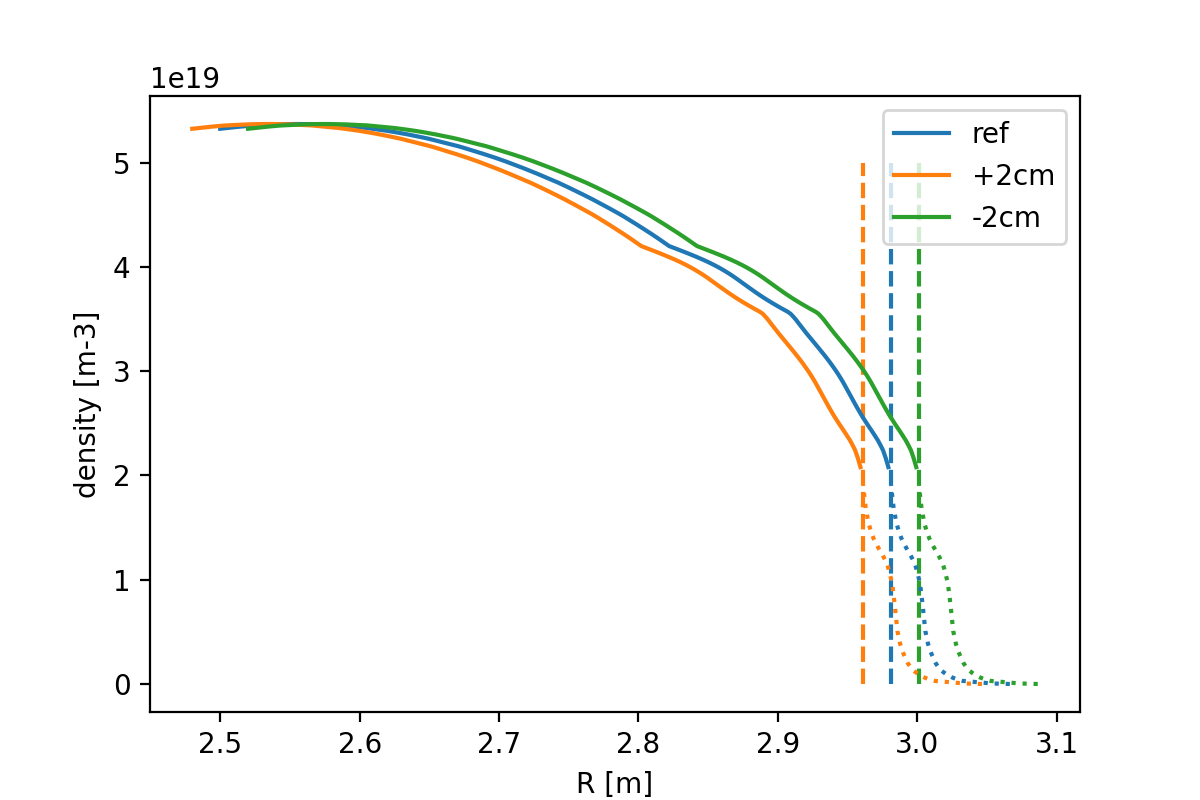}
\end{minipage}
\caption{(a) Shifting the density profile and keeping the LCFS distance cut to the antenna. (b) Shifting the entire profile from the antenna. The second option was used to test the TWA sensitivity to load variations.}
\label{fig:profile_shift}
\end{figure*}

\begin{figure}
    \centering
    \includegraphics[width=\linewidth]{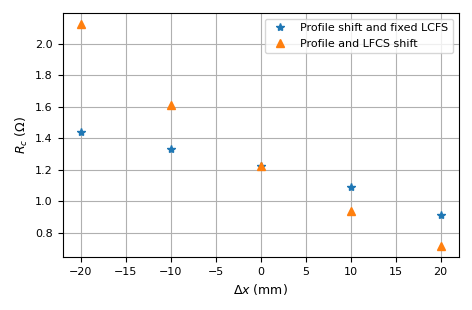}
    \caption{Q2 radiation resistance as a function of the load distance to the antenna. Shifting the density profile and keeping the LCFS distance cut to the antenna or shifting the entire profile from the antenna.}
    \label{fig:Rc_VS_distance}
\end{figure}

\begin{figure}
    \centering
    \includegraphics[width=\linewidth]{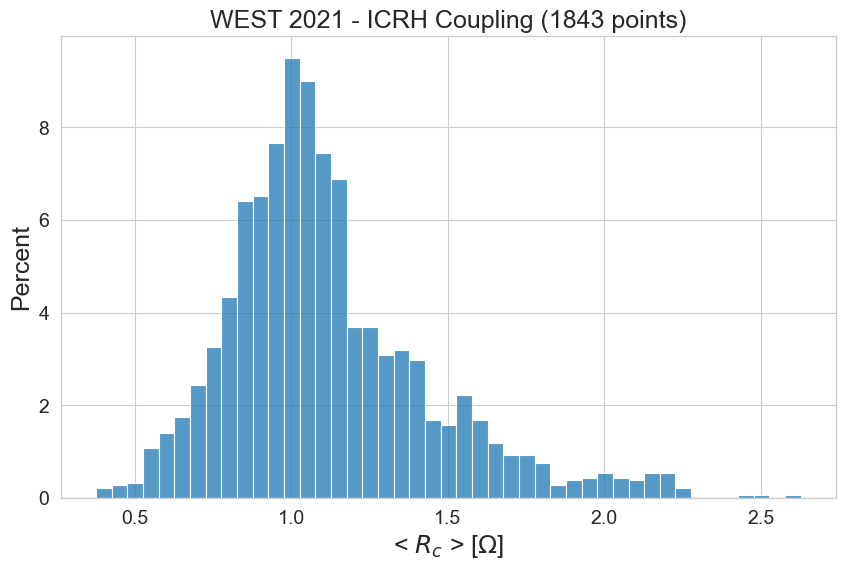}
    \caption{Frequency of radiation resistance $R_c$ measured in the 2021 C5 campaign of WEST.}
    \label{fig:C5_radiation_resistance}
\end{figure}

An important point of the coming analysis is to assess the sensitivity of the fields and the power spectrum of the antenna to a change in plasma loading. To have a reference case, the loading in front of the Q2 antenna was varied in two different ways: 
\begin{enumerate}
    \item shifting the dielectric profile while keeping the LCFS cut at the same position,
    \item shifting the complete profile changing the distance between the antenna aperture and the profile's LCFS. 
\end{enumerate}
The two methods are illustrated in figure \ref{fig:profile_shift}. The range of radiation resistance computed for the first method is 1.4--0.9 $\Omega$ while the second method leads to a larger range 2.1--0.7 $\Omega$ as shown in figure \ref{fig:Rc_VS_distance}. In the C5 WEST campaign, the radiation resistance measured spanned a large range of values, as presented in figure \ref{fig:C5_radiation_resistance}. Therefore, to characterize the antenna sensitivity to load variation, the second option of varying the distance between the antenna and the LCFS was chosen for conservative reasons.

\subsection{HFSS Power spectrum and near fields}\label{subsec:HFSS_power_near fields} \ \\
The analysis of the field and the power spectrum launched by the TWA is performed using the reference dielectric profile described above. The choice of this particular loading in HFSS does not hamper the main conclusions of the following sections.

The fields around the antenna, especially on PFCs like the antenna limiters, should be minimized as well as the low toroidal part of the power spectrum. The minimization of the parallel electric field and the parallel currents on the limiters are of particular concern as they are expected to give rise to local RF sheaths. Their minimization led to the successful operation of ICRH in the W environment of AUG \cite{Bobkov_2016}. Other components of the electric field should also be minimized to reduce ponderomotive effects and other non-linear behavior expected to take place near an ICRH antenna \cite{van2015crude}.

By design, the TWA is already reducing the average field amplitude near the antenna by using a larger number of straps \cite{Ragona_2016} compared to the conventional antennas. These fields on PFCs can be further reduced by changing the capacitor layout and therefore changing the antenna current distribution over the structure. The case of a TWA with and without a toroidally aligned FS was analyzed. The FS bars radial dimensions, orientation and spacing are identical to the Q2 antenna ones.

The power spectrum and the fields excited near the antenna are computed on a surface covering the whole antenna, located 3 mm in front of the limiters. In this paper, the direction parallel to the static magnetic field $B_0$ was taken to be in the toroidal $z$ direction. This is an approximation as the WEST magnetic field is tilted by an approximate angle $\alpha\approx5^\circ$ compared to the machine toroidal direction. A full treatment would take into account the mixing of field components like $E_\parallel=E_z\cos(\alpha)+E_y\sin(\alpha)$. This could lead to wrong results as $E_y$ is significantly larger than $E_z$ near the antenna aperture. However, due to the small $E_y$ field component present near the limiters due to the PEC boundary conditions and the low value of $\alpha$, the minimization of $E_z$ and $E_\parallel$ lead practically to the same results. Moreover, this field correction is expected to be smaller than the correction arising by switching from a flat to a curved antenna model.

Different fields were characterized near the antenna:
\begin{enumerate}
    \item The maximum and average parallel electric fields are computed on a surface 3 mm in front of the limiters. This should be easily reproduced in future TOPICA simulations. Here the toroidal component of the electric field $E_z$ is used as the correction brought by the tilting angle $\alpha$ is small.
    \item The maximum and average parallel current $J_z$ distribution are computed on the limiters. As HFSS solves for the E-field, the B-field is calculated from the E-field. Subsequently, the current density is calculated from the magnetic field. For our analysis, the current density and the electric field were observed to carry the same information. Therefore, only the $E_z$ field results are  presented in the rest of the manuscript.
\end{enumerate}
The goal is to find a minimum of near fields and low-$\abs{k_\parallel}$ power spectrum with a reasonable antenna bandwidth around 53 MHz. This problem proved to be impractical to solve by probing the full 4D space of solutions. The objective was achieved using a minimization with random initial capacitance parameters. This method has the advantage of finding several local minima of antenna near-fields. The best engineering compromise between the near-fields and the antenna power spectra is then selected from the restrained set of solutions found. 

For the fixed geometry defined in figure \ref{fig:TWA_sketch}, a compromise between low fields, power spectrum, and acceptable frequency bandwidth was found for the capacitor combination $(C_1,C_2,C_3,C_4)=(135, 125, 70, 105)$ pF for both the screened and unscreened TWAs. This compromise will be further discussed in the next section and compared with a flat model of the Q2 antenna already installed in WEST. 

\begin{figure}
    \centering
    \includegraphics[width=\linewidth]{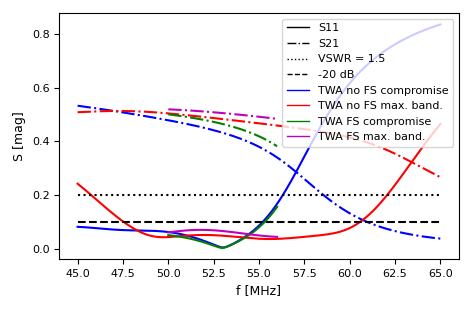}
    \caption{TWA frequency responses for the two capacitor layouts: a TWA tuned to have the maximum frequency bandwidth around 53 MHz or the engineering compromise reducing the antenna fields and power spectrum.}
    \label{fig:TWA_bandwidth}
\end{figure}

This layout can be compared to a typical layout used to maximize the antenna frequency bandwidth \textit{e.g.} $(C_1,C_2,C_3,C_4)=(102,  100,  80,  92)$ pF. The frequency responses of the two antenna cases are presented in figure \ref{fig:TWA_bandwidth}. One can note the close behaviour found between \ATT{} and HFSS. This suggests the possibility to perform a rapid optimization of the antenna geometry in \ATT{}, directly assessing the antenna spectrum and bandwidth, and validating it subsequently with HFSS. 
\newline

From these results, several points can already be underlined:
\begin{itemize}
    \item The frequency bandwidth between the screened and unscreened TWAs is not changing. This is due to the fact that the straps are positioned far from the FS. 
    
    \item The broad minimum found for the average electric field, currents and low-$\abs{k_z}$ power spectrum were coinciding. However, the exact minima found for the average and maximum $E_z$ fields as well as the low-$\abs{k_z}$ ratio of the antenna power spectrum can slightly depart from one another. The frequency response of the antenna is the most limiting factor when finding a compromise between these parameters.
    
    \item For the same amount of coupled power, the screened TWA antenna leads to similar average and maximum $E_\parallel$ on PFCs, and similar ratios of low-$\abs{k_z}$ than the unshielded case. This is not the case for the Q2 antenna, showing larger fields on the limiters in the unshielded case.
\end{itemize}

\subsection{Comparison with WEST Q2 antenna}\label{subsec:comparison} \ 
In order to have a comparison point, the flat models of (i) a shielded TWA, (ii) an unshielded TWA, (iii) a shielded Q2 antenna and (iv) an unshielded Q2 antenna are compared in front of the same load and for a fixed coupled power of 1 W. The two capacitor layouts previously presented are used, \ie: (a) the TWA tuned to have a maximum frequency bandwidth around 53 MHz or (b) the engineering compromise, discussed in section \ref{subsec:HFSS_power_near fields}. 

\begin{figure*}[htb]
    \centering
    \includegraphics[width=\linewidth]{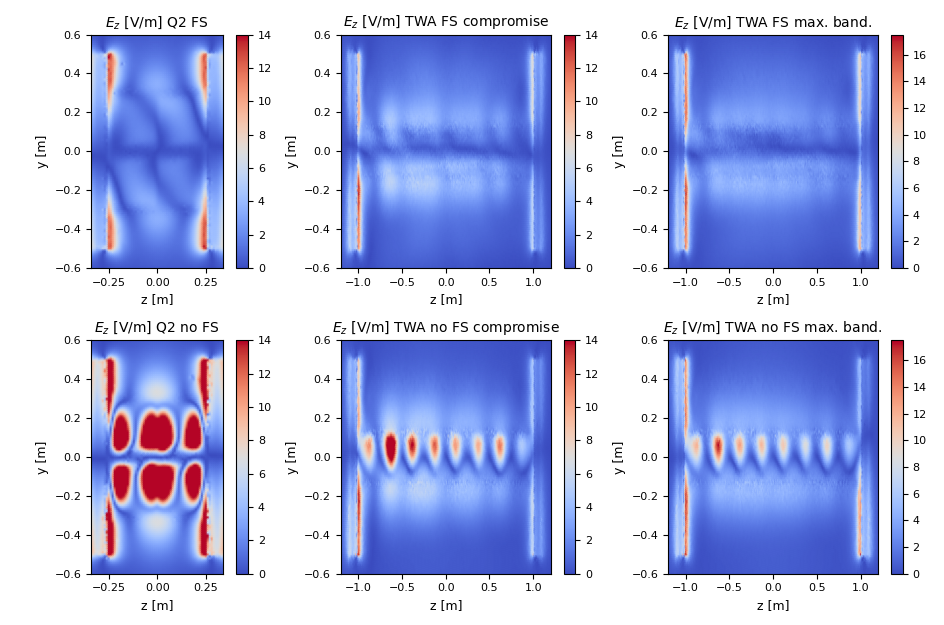}
    \caption{$E_z$ field distribution over a surface 3 mm from the limiters for the shielded and unshielded Q2 and TWA antenna in the case of a tuning maximizing the frequency response of the antenna and one minimizing the fields on the limiters.}
    \label{fig:Ez_distribution}
\end{figure*}

One can first look at the distribution of the electric field $E_z$ on a surface 3 mm from the antenna limiters presented in figure \ref{fig:Ez_distribution}. While the maximum fields taking place on the limiters of the TWA compromise case is either lower or comparable to the Q2 antenna, this maximum only takes place on one limiter and for a \textit{strap recess made 2 cm larger than the Q2 antenna}. This leads to \textit{lower average field} over the two antenna limiters.

\begin{figure}
    \centering
    \includegraphics[width=\linewidth]{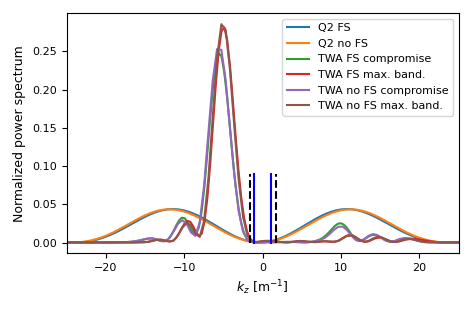}
    \caption{Normalized power spectra for the Q2 WEST antenna and for the screened and unscreened TWA in the case of a tuning maximizing the frequency response of the TWA and one minimizing the fields on the limiters.}
    \label{fig:WEST_vs_TWA_spct}
\end{figure}

The power spectrum launched in the different cases is presented in figure \ref{fig:WEST_vs_TWA_spct}. One observes that the engineering compromise is reducing the ratio of low-$\abs{k_z}$ power spectrum found with HFSS and leads to an optimized power spectrum similar to the one previously computed in \ATT{}. This figure also illustrates the narrow selective spectra obtained with the TWA compared to the Q2 antenna which should be beneficial for power deposition inside the plasma core.

\begin{figure}
    \centering
    \includegraphics[width=\linewidth]{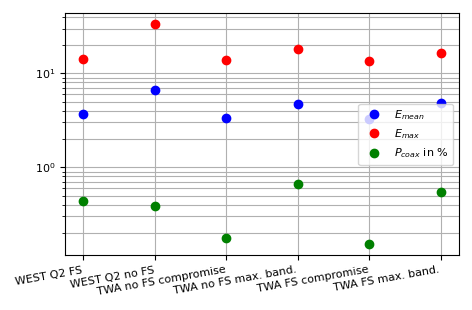}
    \caption{Comparison of $E_{z,mean}$, $E_{z,max}$ (V/m) and $\rho_{k_z<1.5k_0}$ (\%).}
    \label{fig:Q2vsTWA}
\end{figure}

The values of the main parameters of the antenna are summarized in figure \ref{fig:Q2vsTWA}. Going from the WEST Q2 antenna to the unshielded TWA would reduce $E_{z,mean}$ and $E_{z,max}$ on the limiters and the ratio of low-$k_z$ power $\rho_{k_z<1.5k_0}$. One also observes that going from a Q2 antenna with a Faraday screen to one without clearly enhances the $E_z$ fields on the limiters. This does not appear in the TWA case. Going from a Q2 antenna to the case of a TWA tuned to maximize the frequency response leads to higher fields and higher $\rho$ ratios. It is important to note that while the TWA is using a larger number of straps compared to the Q2 antenna, these results are derived \textit{using a strap recess 2 cm larger} than the Q2 antenna. The possibility of only using a 5 cm strap recess instead of the 6 cm presented is presently considered.

\subsection{Sensitivity to capacitance variations} \ 
In this section, the sensitivity of the power spectrum and the near fields to possible capacitance variation is studied. This analysis illustrates the maximum effect of uncertainties arising during antenna manufacturing and from thermal effects arising during operation. The antenna bandwidth is expected to be resilient to these changes as demonstrated in \cite{Ragona_2022}. The analysis is performed for the unshielded TWA but the observations made are also relevant for the shielded case.

\begin{figure}
    \centering
    \includegraphics[width=\linewidth]{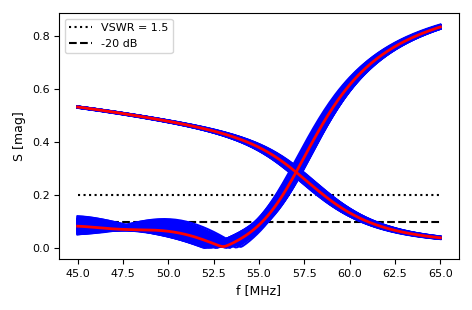}
    \caption{Frequency response variation of the TWA under a random change of capacitance ranging from $0$ to $2$ pF imposed a hundred times on the 7 capacitors tuning the antenna array. The red line represents the reference case.}
    \label{fig:sensitivity_capa_response}
\end{figure}

\begin{figure*}
\begin{minipage}[t]{0.49\linewidth}
    \centering
    \includegraphics[width=\linewidth]{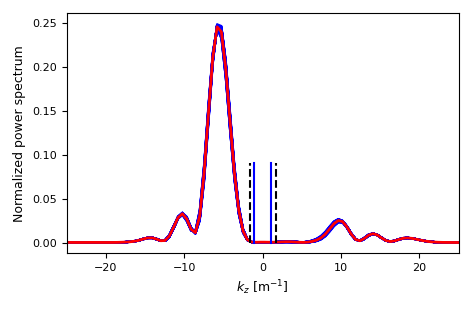}
\end{minipage}
\hfill
\begin{minipage}[t]{0.49\linewidth}
    \centering
    \includegraphics[width=\linewidth]{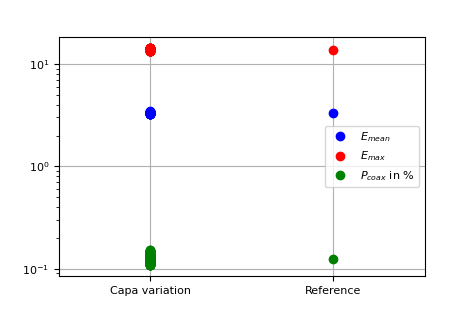}
\end{minipage}
\caption{(a) Spectrum variation under the change of capacitance of figure \ref{fig:sensitivity_capa_response}. (b) Fields and $\rho_{k_z<1.5k_0}$ variation under the change of capacitance of figure \ref{fig:sensitivity_capa_response}. }
\label{fig:sensitivity_capa}
\end{figure*}

To assess this sensitivity, a random change of capacitance ranging from $0$ to $2$ pF was imposed a hundred times on the 7 capacitors tuning the antenna array. The effect of these random errors on the antenna frequency bandwidth, its fields and its power spectrum is presented in figure \ref{fig:sensitivity_capa_response} and \ref{fig:sensitivity_capa}. While the bandwidth of the antenna is slightly impacted, the antenna power spectrum and the maximum and average fields on the limiters remain practically unchanged.

\subsection{Sensitivity to load variations} \
In this section, the sensitivity of the power spectrum and near fields to load variations is studied. This analysis assesses the effect of a change of antenna loading due to the wide range of plasma characteristics found in WEST. The analysis is performed for the unshielded TWA but the observations made are also verified for the shielded case.

\begin{figure}
    \centering
    \includegraphics[width=\linewidth]{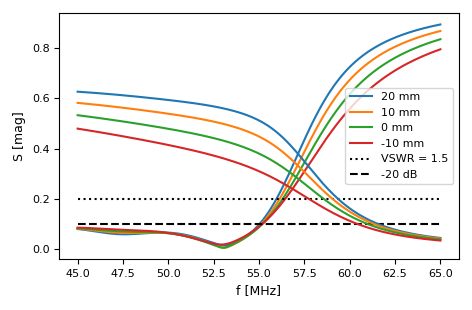}
    \caption{Frequency response variation of the TWA for a shift of the full profile from -10 to 20 mm away from the antenna.}
    \label{fig:sensitivity_load_BW}
\end{figure}

\begin{figure*}
\begin{minipage}[t]{0.49\linewidth}
    \centering
    \includegraphics[width=\linewidth]{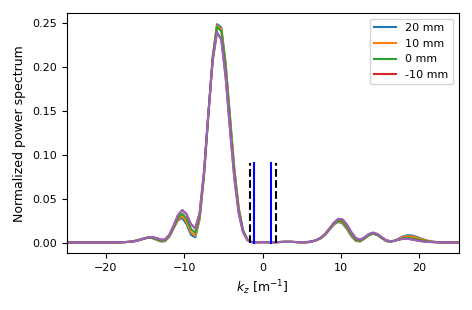}
\end{minipage}
\hfill
\begin{minipage}[t]{0.49\linewidth}
    \centering
    \includegraphics[width=\linewidth]{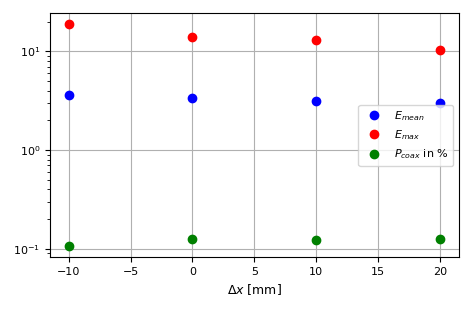}
\end{minipage}
\caption{(a) Spectrum variation under the change of loading of figure \ref{fig:sensitivity_load_BW}. (b) Fields and $\rho_{k_z<1.5k_0}$ variation under the change of loading of figure \ref{fig:sensitivity_load_BW}. }
\label{fig:sensitivity_load_fields}
\end{figure*}

The load imposed in front of the antenna is shifted by a distance $\Delta x$ away from the antenna varying from -10 to 20 mm. The -20 mm case is discarded as it nearly superimposes the equivalent dielectric load boundary with the surface computing the fields. In our Q2 antenna model, this would correspond to a conservative change of radiation resistance of 1.6 to 0.7 $\Omega$. The effect of these variations on the antenna frequency bandwidth is presented in figure \ref{fig:sensitivity_load_BW}. The fields and its power spectrum are presented in figure \ref{fig:sensitivity_load_fields}. The effect of loading variation on the power spectrum launched by the antenna and its frequency response is limited. 
\newline

In our simulations, shifting the LCFS closer to the antenna decreased the fields on the strap while the fields on the limiters increased. This is explained by the fact that the radial electric field $E_x$ taking place in the vacuum gap between the antenna and the dielectric profile increases with a smaller vacuum gap. To satisfy the limiter electric boundary condition, this radial electric field $E_x$ picks up a proportional toroidal component $E_z$. This is further supported by the fact that the same fields decrease when the density profile is shifted closer while keeping the LCFS distance cut to the antenna at a fixed position. The fact that a TWA will excite lower fields due to its larger number of straps compared to a conventional Q2 antenna is still verified by comparing the fields of the TWA and Q2 antenna in front of the same loading as was done in section \ref{subsec:comparison}.

\subsection{Sensitivity to frequency variations} \
This section studies the sensitivity of the antenna power spectrum and fields to the operating frequency. In WEST, the antenna frequency range foreseen spans 46-65 MHz. The analysis is performed for the unshielded TWA but the observations made are also relevant for the shielded case.

\begin{figure*}
\begin{minipage}{0.49\linewidth}
    \centering
    \includegraphics[width=\linewidth]{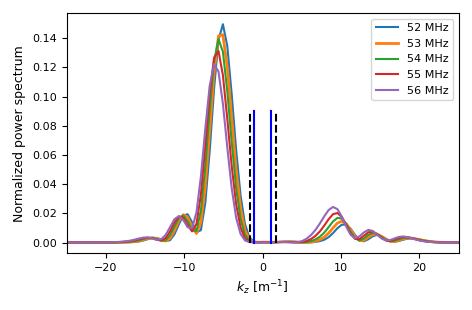}
\end{minipage}
\begin{minipage}{0.49\linewidth}
    \centering
    \includegraphics[width=\linewidth]{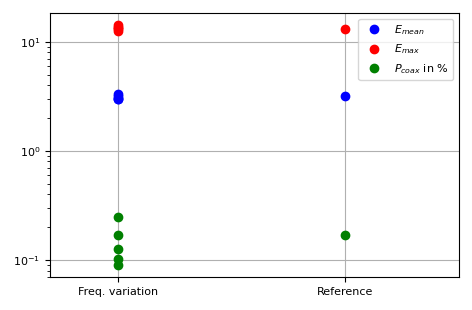}
\end{minipage}
\caption{(a) Spectrum variation under the change of frequencies. (b) Fields and $\rho_{k_z<1.5k_0}$ variation under the same change of frequencies.}
\label{fig:sensitivity_freq}
\end{figure*}

The effect of these variations on the antenna fields and its power spectrum is presented in figure \ref{fig:sensitivity_freq}. The frequency response of the antenna is kept constant and is, therefore, not shown. A change in frequency changes the relative phase between straps. In this configuration, going to larger frequencies leads to lower ratios of low-$\abs{k_z}$ in the power spectrum but only influences slightly the fields, having a greater impact on the maximum field excited. 

\section{Optimization} \label{sec:Optimization}
This section discusses possible amelioration of the antenna model. These optimizations will be further discussed in a future paper which will also discuss the effect of stacking poloidally two TWAs, and the effect of curvature on the results presented above.

\subsection{Antenna optimization}
As can be seen from figure \ref{fig:sensitivity_capa_response}, the antenna response found with the capacitor layout compromise leads to a bandwidth shifted to lower frequencies. Until now, the geometry considered was fixed and the capacitor layout was optimized given this fixed geometry. However, to drive the best engineering compromise, both parameters could be varied at the same time. This could first be tackled in \ATT{} and then validated using HFSS or TOPICA. 

The first geometrical parameter to vary is the inter-strap distance of the antenna. This inter-strap distance largely defines the relative phase difference between the straps and therefore the antenna power spectrum. The strap width could also be varied to change the antenna power spectrum. Another parameter is the gap implemented between the edge straps and the antenna box. This gap could be reduced and would possibly allow a better cancellation of the fields at the antenna limiters provided a new capacitor layout compromise.

\subsection{Antenna coupling}
For the study presented, a strap recess of 6 cm from the antenna aperture was enforced to verify the good antenna coupling brought by the larger number of straps used in a TWA. This leads to TWA radiation resistance similar to the one found for the Q2 antenna which has a strap recess of 4 cm. If the coupling appears as a concern to the WEST team, the strap recess could be reduced to 5 cm. This would lead to a larger coupling and would lead to a larger reduction of fields on the antenna PFCs provided a new capacitor layout compromise.

\section{Conclusions} \label{sec:conclusion}
In this paper, we show the possibility to tailor the current distribution in phase and amplitude in a TWA by changing the capacitor layout used to tune the antenna. This capability gives the possibility to reduce near fields and the low-$\abs{k_z}$ part of the power spectrum launched by the TWA. This fact is demonstrated in front of a plasma load in \ATT{} with a simple antenna geometry and in front of a dielectric load in HFSS with a complex antenna geometry including the limiters. The RF physics presented here is fully applicable to DEMO and brings the TWA a step closer to be a relevant solution for a future ICRF heating system in a future fusion reactor environment like DEMO.

The sensitivity of the near fields and the low-$\abs{k_z}$ ratio to capacitor layout, load and frequency variations is also assessed. As expected from anterior studies and experiments, our study shows that the antenna has a very good resilience to the large variations of loading expected in a reactor environment. The antenna also shows a good resilience to capacitor variations that could arise from manufacturing defaults and thermal effects. Furthermore, a direct tuning of the antenna frequency should be possible and would enable to directly control of the radial power deposition in the plasma.

The paper finally discusses possible second-order optimizations in order to adjust the antenna frequency response, the antenna coupling and the antenna cancellation of near fields on the limiters. 

The next step foreseen for the TWA simulation effort for WEST and DEMO is to validate the results found with a dielectric load in the case of a plasma load with a tool like TOPICA. At the same time, the effect of a poloidal stack of two TWA and the antenna curvature on the physics described above should be characterized. The possibility to include a poloidal phase between a stack of two TWA opens the potential to further minimize fields on the TWA while enhancing its power coupling.

\ack
In memory of A. Messiaen who shared our enthusiasm for TWAs until his last days.
\newline


This work has been carried out within the framework of the EUROfusion Consortium and has received funding from the Euratom research and training programme 2014-2018 and 2019-2020 under grant agreement No 633053. The views and opinions expressed herein do not necessarily reflect those of the European Commission.

\end{document}